\documentclass[fleqn,usenatbib]{mnras}

\usepackage{newtxtext,newtxmath}
\usepackage[T1]{fontenc}
\usepackage{graphicx}
\usepackage{amsmath}
\usepackage{booktabs}

\newcommand{\Lsun}{{\rm L}_{\sun}}
\newcommand{\Tcmb}{T_{\rm CMB}}
\newcommand{\Ltech}{L_{\rm tech}}

\title[Slysh haloes]{Slysh haloes: the waste heat of cold computing as a submillimetre technosignature}

\author[M. A. Garrett]{
M. A. Garrett$^{1,2,3}$\thanks{E-mail: michael.garrett@manchester.ac.uk}
\\
$^{1}$Jodrell Bank Centre for Astrophysics, Department of Physics and Astronomy, The University of Manchester, Manchester M13 9PL, UK\\
$^{2}$Leiden Observatory, Leiden University, PO Box 9513, 2300 RA Leiden, The Netherlands\\
$^{3}$University of Malta, Institute of Space Sciences and Astronomy, Msida, MSD2080, Malta
}

\date{Accepted XXX. Received YYY; in original form ZZZ}
\pubyear{2026}

\begin{document}
\label{firstpage}
\pagerange{\pageref{firstpage}--\pageref{lastpage}}
\maketitle

\begin{abstract} Searches for Dysonian waste heat have operated almost exclusively in the mid-infrared and are therefore sensitive primarily to technology radiating at $\sim$100--600~K. We argue that mature, computation-dominated civilisations may instead dissipate much of their energy at far lower temperatures. The Landauer cost of irreversible computation scales linearly with temperature, while ambient temperatures at large circumstellar radii approach the 2.7~K cosmic microwave background floor. Cold computation is therefore thermodynamically attractive and, because the required radiating area scales as $T^{-4}$, potentially conspicuous. These considerations predict a new object class, which we term the \emph{Slysh halo} after the first advocate of cold Dysonian searches \citep{slysh85}. A Slysh halo is a physically motivated partial ($f\ll1$) Dyson swarm producing grey ($\beta\approx0$), line-free thermal emission from the cold outer regions of planetary systems. We show that such structures are energetically and materially plausible, and that M dwarfs provide especially favourable search targets. Archival far-infrared and submillimetre surveys of nearby stars (DEBRIS, DUNES and SONS) can in principle be reinterpreted to constrain cold circumstellar dissipation at approximately the $10^{20}$~W level ($f\sim10^{-7}$--$10^{-4}$), several orders of magnitude below the waste-heat luminosities targeted by previous infrared searches. Additional opportunities are provided by archival observations from \textit{Planck} and the JCMT, together with the reprocessing of interferometric data from facilities such as ALMA and NOEMA. We assemble six observational discriminants that separate engineered radiators from natural cold sources and outline a three-tier search programme. Even a null result would provide the first temperature-complete assessment of Dysonian technosignatures. 
\end{abstract}


\section{Introduction}
\label{sec:intro}

\citet{dyson60} observed that a civilisation using a significant fraction of
its star's luminosity cannot evade the second law of thermodynamics. Only two
channels could keep the absorbed energy from re-emerging as heat: storing it,
or exporting it from the system as collimated, low-entropy radiation. Neither
can absorb a stellar luminosity indefinitely. Storage capacity is finite, 
chemical media saturate at $\sim$1~eV per atom, and a structure that
accumulates energy at stellar rates soon becomes chemically and
gravitationally unbound \citep{lacki16,wright20}.  What remains
in practice, is re-emitted as thermal waste heat, and
the resulting infrared excess is the archetypal technosignature. 

Sixty years of observational work has followed: IRAS-based searches \citep{slysh85,timofeev00,jugaku04,carrigan09}, WISE-based programmes on a galactic scale \citep{wright14,griffith15,garrett15,chen21}, and most recently the Project Hephaistos \textit{Gaia}+2MASS+WISE photometric survey of five million stars \citep{suazo24}. Follow-up radio, mid-infrared and near-infrared observations of the Hephaistos candidates suggest that many are contaminated by background dust-obscured galaxies \citep{ren24,ren25,ren26,zackrisson26}. These programmes share a rarely examined restriction: their wavebands confine
them to structures radiating at roughly 100--600~K -- the temperatures of
habitable-zone engineering. 

This paper develops the opposite limit. Its premise, set out in
Section~\ref{sec:thermo}, is that the energy budgets of mature technological
civilisations are plausibly dominated by computation
\citep{cirkovic06,sandberg16,garrett24, 2026PASP}, and that computation, unlike habitation or industry, has a strong, quantifiable thermodynamic
preference for operating cold: the Landauer bound scales linearly with
temperature \citep{landauer61}, refrigeration below ambient incurs a
significant Carnot penalty while ambient cooling at large circumstellar
radius is free,
and superconducting hardware becomes passively available beyond a few tens of
au. The same logic that led \citet{cirkovic06} to predict migration of
post-biological intelligence toward the cold outer Galaxy also operates within each
planetary system, and drives mature infrastructure outward and down the
temperature ladder toward the hard floor set by the cosmic microwave
background \citep[CMB; $\Tcmb = 2.7255$~K,][]{fixsen96}. Nor is the premise
merely speculative. Within a century of building its first computers,
humanity's fastest-growing demand for energy is computation, and proposals to
move that computation off-planet -- constellations of solar-powered,
optically interconnected orbital data centres
\citep[e.g.\ Project Suncatcher;][]{suncatcher25,marcy26} --
have entered engineering development, motivated by exactly the
considerations above: uninterrupted sunlight and passive radiative cooling to
cold sky. The migration this paper
extrapolates towards has arguably already begun. Nor need the subsequent climb be
slow - simple growth models show that at AI-era demand growth rates, the
transition from planetary to stellar-scale energy capture compresses from
Kardashev's millennia into centuries \citep{garrett26a,nachtrieb26}.

The observational consequence is a new, predicted object class, which we propose to
call the \textbf{Slysh halo}: a population of computing structures
distributed through the region of a planetary system where the ambient
temperature falls to $\simeq$28--5~K. This corresponds to distances of tens of au for the M dwarfs that
dominate the stellar population, and $\sim$100--3000~au for a star like the
Sun. Slysh haloes are powered by collected starlight (Section~\ref{sec:energetics}), built from the small-body
material that formation processes leave at such radii
(Section~\ref{sec:mass}), and radiate their entire power budget as grey
thermal emission at the local equilibrium temperature,
$T(r) \simeq 278.3\,(r/{\rm au})^{-1/2}$~K -- the standard blackbody
equilibrium law, derived in Section~\ref{sec:whycold}
\citep[e.g.][]{wyatt08} -- i.e.\ at $\simeq$28--5~K, peaking in the
far-infrared and submillimetre. In the language of Dysonian SETI the halo is a
partial swarm with covering fraction $f \ll 1$, but one whose location,
temperature and spectrum follow directly from thermodynamics. The name
honours V.~I. Slysh, who first argued that thermodynamic efficiency drives
astroengineering toward low temperatures and first proposed the
corresponding infrared-to-microwave search \citep{slysh85}. The
host star remains optically normal. The halo shares its parallax and proper
motion. And each component is faint where the other is bright: the star is
reduced to a weak Rayleigh--Jeans tail at submillimetre wavelengths, while
the halo emits nothing in the optical.

\citet{slysh85} supplied the template
the current paper resumes: arguing from Carnot efficiency,
$\eta = 1 - T_{\rm DS}/T_{*}$, that engineered spheres should run cold and
identifying the 2.7~K background as the ultimate floor. Slysh recognised that these systems would appear as far-infrared-to-millimetre
sources "without any optical counterpart", and noted the red-giant
confusion problem in the IRAS data of the day. \citet{timofeev00} fitted $3<T<300$~K Planck spectra to IRAS sources and \citet{lacki16} argued from Landauer's principle that
computation-maximising societies should operate just above the CMB, searching the \textit{Planck} compact-source catalogue
for galaxy-spanning examples. \citet{wright20} and \citet{wright23} have supplied the modern formal treatments of Dysonian SETI, the latter deriving from the thermodynamics of radiation, the
efficiency and computation-rate limits on which the argument of this paper
rests (Section~\ref{sec:whycold}).

 What does not exist in the literature, to our knowledge,
is (i) the circumstellar halo as a predicted object class with its energetic
and material budgets worked out; (ii) the recognition that the modern
submillimetre sky surveys already reach the required depths
around nearby stars to make interesting searches possible; (iii) a discriminant suite against the natural cold
confounding populations; and (iv) the reinterpretation of archival debris-disc photometry
as first order limits on engineered circumstellar structures.  

Supplying these is the purpose of
this paper: to revive Slysh-style cold Dysonian SETI at the
stellar scale, and confronting it with data that did not exist when it was originally proposed. Section~\ref{sec:thermo} presents the thermodynamic argument;
Section~\ref{sec:halo} defines the Slysh halo and its energetic and
material requirements; Section~\ref{sec:model} sets out the grey-body
observational model used throughout; Section~\ref{sec:observability} derives
the halo's observational signatures; Section~\ref{sec:limits} extracts limits
from archival data for partial swarms and treats the complete-shell limiting
case; Section~\ref{sec:discriminants} assembles the discriminants and
confusion budget; Section~\ref{sec:programme} outlines a possible tiered search observing
programme; and Section~\ref{sec:discussion} discusses the wider implications of this work.

\section{The thermodynamics of cold computation}
\label{sec:thermo}

\subsection{Why mature technology computes cold}
\label{sec:whycold}

Three independent arguments favour low-temperature information processing.

\textit{(i) The Landauer bound.} The minimum energy dissipated per
irreversible bit operation is
\begin{equation}
E_{\rm L} = k_{\rm B}T\ln 2
\simeq 9.57\times10^{-23}\left(\frac{T}{10\,{\rm K}}\right)~{\rm J}
\label{eq:landauer}
\end{equation}
\citep{landauer61}: $2.9\times10^{-21}$~J at 300~K but $9.6\times10^{-23}$~J
at 10~K. Any technology operating near the bound performs, per joule, thirty
times more irreversible operations at 10~K than at room temperature. The
choice of operating temperature is fundamentally a choice between optimising
for computational speed and computational efficiency: the Margolus--Levitin
theorem \citep{margolus98} -- which limits any physical system of mean energy
$E$ to at most $4E/h$ orthogonal state transitions per second -- ties maximum
processing \emph{rate} to energy, but the total number of irreversible
operations a society can perform over its lifetime is bound by the Landauer
limit, and for a fixed energy budget cold operation wins.  Applying the formalism of \citet{landsberg80}, \citet{wright23} showed that for a structure that intercepts
luminosity $L$ and re-radiates at temperature $T$, the maximum number of irreversible operations per second is given by
\begin{equation}
r = \frac{4}{3}\,\frac{L}{k_{\rm B}T\ln 2}\left(1-\frac{T}{T_{*}}\right)
\label{eq:wrightrate}
\end{equation}
(\citet{wright23} equations 23 and 27): the Carnot factor, already written down in this context by \citet{slysh85}, arises because work is extracted
between the stellar radiation temperature $T_{*}$ and the radiator
temperature, and the factor $4/3$ because blackbody radiation carries entropy
$\frac{4}{3}\sigma T^{3}$ per unit area, making bulk radiative entropy
disposal slightly cheaper than the differential Landauer figure suggests. At
fixed $L$, equation~(\ref{eq:wrightrate}) rises monotonically -- approximately
as $1/T$ -- as the radiator
cools. The plain Landauer accounting used in this paper is therefore
conservative, and the thirty-fold advantage of 10~K over 300~K operation
carries over essentially unchanged. 
\citet{lacki16} invoked the same principle on galactic scales,
arguing that a computation-maximising society converges on structures at
temperatures just above the CMB; the halo of Section~\ref{sec:halo} is the
circumstellar application of that logic.

\textit{(ii) Free ambient cooling.} Operating below ambient temperature
requires refrigeration: removing heat $Q$ from hardware at temperature $T$
and rejecting it to surroundings at $T_{\rm amb}$ costs work
$W \geq Q\,(T_{\rm amb}-T)/T$ -- the ideal-refrigerator (Carnot) limit, a
direct consequence of the second law -- and the penalty per unit heat
diverges as $T \rightarrow 0$. The penalty
vanishes for infrastructure operating \emph{at} ambient -- and low ambient is
freely available at large circumstellar radius. Consider a body of radius $s$
at distance $r$ from a star of luminosity $L_{*}$. It intercepts starlight
over its cross-section $\pi s^{2}$ and re-radiates thermally over its full
surface $4\pi s^{2}$. If the stellar-band absorptivity $\alpha_{*}$ and
the far-infrared emissivity $\epsilon_{\rm IR}$ are equal, they cancel from
the radiative balance,
\begin{equation}
\frac{L_{*}}{4\pi r^{2}}\,\pi s^{2} = 4\pi s^{2}\,\sigma T_{\rm eq}^{4},
\label{eq:balance}
\end{equation}
which fixes the equilibrium temperature independently of the body's size:
\begin{equation}
T_{\rm eq}(r) = \left(\frac{L_{*}}{16\pi\sigma r^{2}}\right)^{1/4}
\simeq 278.3 \left(\frac{L_{*}}{\Lsun}\right)^{1/4}
\left(\frac{r}{\rm au}\right)^{-1/2}~{\rm K},
\label{eq:teq}
\end{equation}
the standard blackbody equilibrium temperature familiar from debris-disc
studies \citep[e.g.][]{wyatt08}. For a solar-luminosity host this gives 28~K
at 100~au and 12~K at 550~au, approaching $\Tcmb$ (the floor below which no
radiator can reject heat) beyond $\sim$$10^{4}$~au. The whole scale
contracts as $L_{*}^{1/2}$ for fainter hosts: around an M5 dwarf of
$0.0015\,\Lsun$ the same two temperatures occur at 3.8 and 21~au and the
floor is reached near 400~au, so the entire usable range fits within a region
the size of our own planetary system (Section~\ref{sec:hostL}). Within the Galactic disc the practical floor sits a little above this cosmological one: the local interstellar radiation field carries $\sim$0.5--1~eV\,cm$^{-3}$ against the CMB's 0.26~eV\,cm$^{-3}$, so a grey
absorber settles near 3.5--4~K, the exact value depending on the specific Galactic
environment.

\textit{(iii) Passive superconductivity.} Ambient temperature falls below the
transition temperatures of high-$T_{\rm c}$ superconductors ($\sim$93~K)
beyond $\sim$9~au of a solar-luminosity star, MgB$_2$ (39~K) beyond
$\sim$51~au, and elemental niobium (9.3~K) beyond $\sim$900~au, making
dissipationless interconnects, low-thermal-noise detectors and
quantum-coherent hardware available with no cryogenic overhead. These radii
scale as $L_{*}^{1/2}$ - around an M5 dwarf the same three thresholds fall at
0.35, 2.0 and 35~au, so even niobium-class operation is available at
Kuiper-belt distances rather than far outside the planetary region.

\subsection{Objections}
\label{sec:objections}

Two published counterarguments should be noted. First, \citet{sandberg16}
argue that a computation-maximising civilisation optimises by
\emph{aestivating} (the hot weather equivalent of hibernation) - lying dormant through the present, comparatively warm
cosmological era, just as animals aestivate through summer. These civilisations wait until the universe is colder and each joule of energy buys more erasures. The searches proposed here
are agnostic on this point: aestivation still requires infrastructure that
monitors, maintains and computes now, and the aestivation optimum has itself
been disputed \citep{bennett19}: the present universe already contains vast
reservoirs far from maximum entropy into which the waste entropy of
computation can be exported, so operations performed now need cost no more
than operations deferred.

Second, \citet{wright23} -- the most complete published treatment of Dyson
spheres as work extractors and computational engines, and the formalism on
which the efficiency statements of this paper rest -- concludes that the
expected waste heat of technology is warm, perhaps warmer than the classical
habitable-zone structure. It is worth being precise about what that analysis
shows, because its conclusions divide cleanly along the constraint assumed.
Absent a mass constraint, its optimum is the cold optimum advocated here: for
a structure processing a given luminosity, ``the most efficient configuration
is one that maximizes $R$ and minimizes $T$'' \citep[section 5.1]{wright23},
with the computation rate of equation~(\ref{eq:wrightrate}) rising
monotonically as the radiator cools. The warm expectation enters entirely
through mass economics: because radiating area scales as $T^{-4}$
(Section~\ref{sec:area}), the computation rate of a complete structure grows
only as $M^{1/4}$, so doubling the output of a shell costs sixteen times its
mass, and for a small fixed collector area the work-optimal placement is
close to the star, essentially as hot as the hardware allows. 
We do not dispute these scalings; we suggest, rather, that mass need not be
the binding constraint in the regime relevant to detection, and that the
alternative deserves observational attention. The small--hot optimum maximises
computation \emph{rate per unit mass}; the cold optimum maximises total
computation \emph{per unit energy}, the relevant figure of merit for an
energy-limited civilisation, and increasingly so for a long-lived one, for
which the integrated Landauer saving dominates construction cost. 
Section~\ref{sec:mass} engages the mass constraint directly and shows that,
at thin-film areal densities, the $T^{-4}$ mass penalty of cold operation
remains modest up to astronomically detectable dissipation levels, and is
paid in small-body debris, the cheapest mass in any planetary system. Which
constraint binds a real civilisation cannot be settled a priori; the two
regimes predict waste heat at opposite ends of the temperature axis, and the
hot end is already well searched. That asymmetry alone motivates the cold
search.

\subsection{The radiator-area theorem: cold means large, not faint}
\label{sec:area}

The property that makes cold infrastructure conspicuous is forced by the
physics that makes it efficient. From the Stefan-Boltzmann law, rejecting power $P$ at temperature $T$
against the CMB requires a radiating area
\begin{equation}
A_{\rm rad} = \frac{P}{2\,\epsilon\,\sigma\,(T^{4}-\Tcmb^{4})}
\label{eq:area}
\end{equation}
for two-sided flat radiators of emissivity $\epsilon$ (for a spherical shell,
$A_{\rm rad}=4\pi R^{2} = P/[\epsilon\sigma(T^{4}-\Tcmb^{4})]$). The factor 2
counts both faces of a thin panel: collection and rejection are separated by
wavelength, not by side. The sunward face absorbs starlight at optical
wavelengths while emitting thermally at hundreds of microns, just as the
anti-sunward face does; and from hundreds of AU the star subtends a
negligible solid angle, so both faces see essentially the 2.7-K sky. The area
scales as $T^{-4}$: $\sim$$10^{-3}$~m$^{2}$\,W$^{-1}$ at 300~K,
11~m$^{2}$\,W$^{-1}$ at 30~K, 425~m$^{2}$\,W$^{-1}$ at 12~K. Cold computation
is therefore necessarily vast in area, and its thermal emission
correspondingly unavoidable: at fixed dissipated power, lowering $T$
redistributes the emission into the submillimetre and spreads it over a larger
surface; it does not diminish it. This invalidates the common intuition that
cold technosignatures must be intrinsically faint. Bolometric detectability is
governed by the dissipated power rather than by the temperature at which it
emerges, and within a given band the temperature enters through
$\Delta B_{\nu}(T)/(T^{4}-\Tcmb^{4})$ (see equation~\ref{eq:snu}) in a sense that
rewards cold operation: at fixed power a 10-K halo is fourteen times brighter
at 345~GHz than a 30-K one (Table~\ref{tab:fluxes}).

\section{The Slysh halo}
\label{sec:halo}

\subsection{Definition and structure}
\label{sec:definition}

We define a Slysh halo as a distributed population of energy-dissipating
structures occupying the region of a planetary system where ambient
temperature is low enough for efficient computation but the host star's
resources still remain accessible. 
The halo is defined by temperature rather than distance from the star. We take the relevant range in ambient equilibrium temperature to be roughly 28 to 5~K. By equation~(\ref{eq:teq}), the radii corresponding to this temperature range scale as $L_*^{1/2}$. This equates to halo radii of tens of au around the nearby M dwarfs that dominate the solar neighbourhood, and approximately 100--3000~au for a solar-luminosity star (Section~\ref{sec:hostL}). Unless stated otherwise, numerical examples refer to a solar-luminosity
host. Individual nodes of the halo are clearly both 
unresolvable and  undetectable at interstellar distances (a
$10^{12}$~W node at 20~pc presents only $\sim$$10^{-2}$~nJy at 345~GHz;
equation~\ref{eq:snu} with $fL_{*} \rightarrow 10^{12}$~W and $T = 12$~K). The practical 
observable is the \emph{aggregate} dissipated power $\Ltech$, radiated
as grey thermal emission at the local equilibrium temperature given by
equation~(\ref{eq:teq}). Because the nodes orbit the star, the halo shares the
stellar parallax and proper motion; and because it need not intercept the
stellar light cylinder at small radii, the star remains optically
unremarkable. There is no requirement of, and no plausible motivation for, an
enclosing shell; the shell appears in this framework only as the
$f \rightarrow 1$ covering-fraction limit (Section~\ref{sec:shell}). In the
swarm formalism of \citet{wright23} the halo is the low-optical-depth limit,
$\tau = A/4\pi r^{2} \ll 1$, in which self-shadowing among nodes is
negligible and each element radiates freely to space.

Two clarifications on the name. `Halo' is meant in the astronomical sense of
an extended envelope surrounding the star rather than a ring. Because the
structure is optically thin, an external observer sees emission projected
across its entire face, including through the centre where the
stellar host sits, and the surface-brightness profile follows the
radial distribution of nodes together with the temperature law of
equation~(\ref{eq:teq}). What that profile looks like depends on where the
infrastructure sits: nodes crowded inward give a centrally peaked image, while
a geometrically narrow shell is limb-brightened by the longer sightline
through its edge. The profile is thus a measurement of the radial
distribution rather than a fixed prediction, and we return to it in
Section~\ref{sec:observability}. Nor is sphericity
assumed: infrastructure grown from a Kuiper-belt analogue may equally be a
thick disc or torus. The aggregate photometry of Section~\ref{sec:model} is
insensitive to this geometry for isotropically oriented panels, the
observables being $\Ltech$ and $T$ alone; a flattened disc or torus of
star-facing panels carries in addition an inclination dependence through its
projected area, and
morphology enters only once the halo is resolved
(Section~\ref{sec:observability}).

\subsection{Energetics: the collector--radiator identity}
\label{sec:energetics}

An immediate objection is that starlight at large distances is too feeble to power significant computation. The
local irradiance $F(r) = L_{*}/4\pi r^{2}$ is only
$\sim$$5\times10^{-3}$~W\,m$^{-2}$ at 550~AU of a solar-luminosity host for example. The objection is addressed 
by noting that the ambient temperature is set by that same flux. A flat
sheet absorbing on its sunward face and radiating from both settles at the
temperature $T_{\rm amb}$ for which
$F(r) = 2\sigma(T_{\rm amb}^{4}-\Tcmb^{4})$, a factor $2^{1/4}$ above
equation~(\ref{eq:teq}): the sheet radiates over twice its collecting area
where a sphere radiates over four times, so it must run hotter to shed the
same absorbed flux.

Consider a node built from such a sheet. Over a collector area $A_{\rm c}$, it intercepts a stellar power of $F(r)\,A_{\rm c}$. Since all collected energy is ultimately dissipated as heat (Section~\ref{sec:intro}), the node must reject this same power from its radiator area $A_{\rm r}$. Operating at temperature $T$, the radiator sheds $2\sigma(T^{4}-\Tcmb^{4})$ per unit area.

Steady-state energy balance therefore requires:
\begin{equation}
F(r)\,A_{\rm c} = 2\sigma(T^{4}-\Tcmb^{4})\,A_{\rm r}
\end{equation}

Rearranging this gives the ratio of required collector area to radiator area:
\begin{equation}
\frac{A_{\rm c}}{A_{\rm r}} = \frac{2\sigma\left(T^{4}-\Tcmb^{4}\right)}{F(r)}
\end{equation}

If the node operates exactly at the local ambient temperature ($T = T_{\rm amb}$), the numerator is simply the definition of the local ambient flux, $F(r)$. The right-hand side therefore evaluates precisely to unity. In other words, $A_{\rm c} = A_{\rm r}$: the collector area equals the radiator area at any distance. Dilution
imposes no penalty because the radiator requirement of
equation~(\ref{eq:area}) grows with distance at exactly the same rate, and the
collector is the same gossamer panel with a photovoltaic sunward face -- the
satellite element assumed by \citet{wright23}. Radii quoted in this paper use
the spherical convention of equation~(\ref{eq:teq}); the observables $\Ltech$
and $T$ do not depend on the choice.

The thermodynamic quality of the arrangement is high. Work is extracted from
5800~K photons and rejected at $\sim$10~K, a Carnot efficiency
$\eta = 1 - T/T_{*} \ga 0.998$ \citep[][equation 34]{wright23}, and each joule
of that work purchases $\sim$30$\times$ more Landauer erasures than it would
at 300~K.

The identity assumes that each node collects its own starlight, but the halo
signature does not depend on that choice: however the energy arrives, it must
leave as thermal emission at the ambient temperature, so the observable of
Section~\ref{sec:model} is unchanged. 


\subsection{An energy-source diagnostic}
\label{sec:diagnostic}

The energy-source question, unresolvable a priori, becomes an observable
after detection. A starlight-powered halo cannot dissipate more than its star
supplies, $\Ltech = fL_{*} \leq L_{*}$; and for any detected halo $f$ is
directly measurable, because \textit{Gaia} supplies the distance and stellar
luminosity while the submillimetre photometry supplies $\Ltech$
(equation~\ref{eq:snu}). A halo with $\Ltech \leq L_{*}$ is therefore
consistent with living off its star, while $\Ltech > L_{*}$ would establish that
present-day starlight is not the sole supply -- onboard generation, stored
energy, accretion, an unseen companion or power beamed from elsewhere would need to be invoked. 

\subsection{Mass budget}
\label{sec:mass}

The cost of cold operation is radiator area; the cost of area is mass. From equation~(\ref{eq:area}), a halo dissipating $\Ltech$ at temperature $T$ requires a total two-sided film area \[ A = \frac{\Ltech} {2\,\epsilon\,\sigma\left(T^{4}-\Tcmb^{4}\right)}, \] and because the collector and radiator are the same sheet (Section~\ref{sec:energetics}) no additional collecting area is required. 
For film of areal density $\sigma_{\rm a}$ the mass is simply $M=\sigma_{\rm a}A$, giving 
\begin{equation} \begin{split} M &= \frac{\Ltech\,\sigma_{\rm a}} {2\,\epsilon\,\sigma \left(T^{4}-\Tcmb^{4}\right)} \\ & \simeq 2.1\times10^{20}\, \frac{\Ltech}{10^{20}\,{\rm W}}\, \frac{\sigma_{\rm a}}{5\,{\rm g\,m^{-2}}} \left(\frac{T}{12\,{\rm K}}\right)^{-4} ~{\rm kg}. \end{split} \label{eq:mass} \end{equation} Thus the mass depends only on the power to be dissipated and the temperature at which it is rejected, not on the luminosity of the host star. The numerical coefficient assumes a bolometric emissivity $\epsilon=1$; a less emissive surface must deploy proportionately more area to shed the same power, so $M\propto1/\epsilon$. Separate collector and radiator structures would double the mass, as would a filled shell, which radiates from one side only. For most of the temperatures of interest here, $T\gg T_{\rm CMB}$ and the denominator is very nearly $T^{4}$, giving the simple and important scaling $M\propto T^{-4}$. The CMB correction modifies the mass by only 0.3 per cent at 12~K and 9 per cent at 5~K, but becomes increasingly important as $T$ approaches the thermodynamic floor. 
Present-day solar-sail membranes already achieve
$\sigma_{\rm a} \sim 1$--10~g\,m$^{-2}$ \citep{russo2022} and the Breakthrough Starshot project aims for values less than this \citep{parkin2018}. Supporting structures raise the
effective figure, which is one reason equation~(\ref{eq:mass}) keeps
$\sigma_{\rm a}$ explicit. Table~\ref{tab:mass} evaluates
equation~(\ref{eq:mass}) at $\sigma_{\rm a} = 5$~g\,m$^{-2}$ and $T = 12$~K.

\begin{table}
\centering
\caption{The mass ladder for Slysh haloes
($\sigma_{\rm a}=5$~g\,m$^{-2}$, $T=12$~K; equation~\ref{eq:mass}), quoted as
swarms; a filled shell of the same power costs twice as much. Masses depend on
$\Ltech$ and $T$ alone and not on host luminosity, which enters only through
the covering fractions in the first column. ``Detectable'' refers to the
archival limits of Section~\ref{sec:partial}. For reference: comet Halley
$\sim$$2\times10^{14}$~kg; Ceres $\sim$$9\times10^{20}$~kg; present Kuiper
belt $\sim$$0.01\,{\rm M}_{\oplus} \sim 6\times10^{22}$~kg; Earth
$6\times10^{24}$~kg; Jupiter $1.9\times10^{27}$~kg.}
\label{tab:mass}
\setlength{\tabcolsep}{4pt}
\begin{tabular}{lll}
\toprule
$\Ltech$ (W) & $M$ (kg) & Equivalent \\
\midrule
$2\times10^{13}$ (humanity)          & $4\times10^{13}$   & 0.2 Halley \\
$7\times10^{19}$ (detectable, 10 pc) & $1.5\times10^{20}$ & 0.2 Ceres \\
$3.8\times10^{22}$ ($f=10^{-4}$, Sun) & $8\times10^{22}$  & 1.3 Kuiper belts \\
$5.7\times10^{23}$ ($f=1$, M5V)      & $1.2\times10^{24}$ & 0.2 M$_{\oplus}$ \\
$3.8\times10^{26}$ ($f=1$, Sun)      & $8\times10^{26}$   & 0.4 Jupiter \\
\bottomrule
\end{tabular}
\end{table}

Three conclusions follow. Civilisation-scale computing -- our own current
power budget, relocated and run 30$\times$ more efficiently -- costs a fifth
of a comet. Astronomically detectable haloes cost a fraction of a Ceres,
extracted not from a planet but from the small-body reservoirs that planet
formation strands at precisely the halo radii, in bodies with escape
velocities of metres per second. And full stellar-luminosity reprocessing
demands of order half a Jupiter of thin film only around a luminous host:
because the requirement follows the power dissipated rather than the fraction
intercepted, complete capture around an M5 dwarf costs a fifth of an Earth
mass, so the $f \rightarrow 1$ limit is far less remote for the faint stars
that dominate the solar neighbourhood (Section~\ref{sec:hostL}). 
This is also where the
mass-constraint argument of \citet{wright23} is engaged quantitatively: the
$T^{-4}$ penalty is real, but at thin-film densities it remains affordable up
to $\sim$$10^{-4}\,\Lsun$ using material that is, in any case, debris. We
state the scaling explicitly in equation~(\ref{eq:mass}) so that readers may
substitute their own pessimism about $\sigma_{\rm a}$; even at
100~g\,m$^{-2}$ the detectable-halo case costs only a few Ceres.

\subsection{Dynamics and stability}
\label{sec:dynamics}

Infrastructure of this kind has to stay where it is put, and the halo zone is
a benign place to try. Nodes there follow ordinary Keplerian orbits at speeds
of a few kilometres per second -- 1.3~km\,s$^{-1}$ at 550~AU of a
solar-luminosity star, and 2.5~km\,s$^{-1}$ at the 21~AU that corresponds to
the same temperature around an M5 dwarf -- with orbital periods ranging from
decades at the inner edge of an M dwarf's zone to a hundred thousand years at
the outer edge of the Sun's. There is no gas to provide drag and no
atmosphere, and tidal distortion by the star is negligible at these
distances.

The one force that distinguishes gossamer film from ordinary debris is
radiation pressure. Starlight pushes a panel outward while gravity pulls it
in, and because both fall off as $r^{-2}$ their ratio
$\beta = L_{*}/(4\pi G M_{*} c\,\sigma_{\rm a}) \simeq
0.15\,(\sigma_{\rm a}/5~{\rm g\,m^{-2}})^{-1}$ is the same at every distance
from the star. Its effect is simply to weaken gravity: the net central force
becomes $GM_{*}(1-\beta)/r^{2}$, so a node moves exactly as it would around a
star of mass $(1-\beta)M_{*}$. Film at 5~g\,m$^{-2}$ therefore orbits a star
that is effectively 15 per cent lighter than the real one, which alters
orbital speeds and periods but leaves the orbits themselves closed and bound.

Two thresholds are worth knowing. A panel unfurled from a parent body that is
already on a circular orbit keeps the speed it had, but now finds itself in
this weakened gravitational field; if $\beta$ exceeds $1/2$ that speed exceeds
the local escape speed, and the panel leaves the system altogether on a
hyperbolic orbit. This is the standard blow-out condition, and for the
parameters used here it corresponds to film lighter than about
1.5~g\,m$^{-2}$. Only above $\beta = 1$, lighter than about
0.8~g\,m$^{-2}$, does radiation pressure exceed gravity outright, so that no
bound orbit exists at any distance at all. Film at the density assumed in
Section~\ref{sec:mass} is comfortably clear of both limits; anything much
lighter would have to be placed deliberately rather than simply let go.

Collisions and mutual shadowing are not a concern either, because a halo is
extraordinarily sparse. Its covering fraction -- the fraction of the star's
sky that the panels block -- has a simple value. Whatever fraction of the sky
the halo covers is the fraction of the starlight it intercepts, and since
everything intercepted is eventually dissipated, that fraction must equal
$\Ltech/L_{*}$, which is just $f$. The geometry says the same thing: the panel
area needed at temperature $T$ grows as $T^{-4} \propto r^{2}$
(equation~\ref{eq:teq}), at precisely the rate the area of a sphere of radius
$r$ grows, so the covering fraction is identical at every radius in the zone.
For the halo that Table~\ref{tab:mass} lists as detectable at 10~pc --
$3\times10^{22}$~m$^{2}$ of film dissipating $7\times10^{19}$~W -- it is
$2\times10^{-7}$. Seen from the star, the sky is essentially empty. Adjacent
orbits shear past one another at metres per second, and the Galactic tides and
stellar encounters that stir the Oort cloud only become significant beyond
$10^{4}$~AU. Station-keeping demands are correspondingly modest. Whether a
civilisation would choose to manufacture a Ceres of thin film is not something
we can know; what we can say is that nothing in the orbital mechanics forbids
it.

\section{A grey-body observational model}
\label{sec:model}

Throughout this paper we adopt the deliberately conservative model of an
optically thick grey radiator of temperature $T$ and bolometric waste
luminosity $\Ltech = f L_{*}$. For starlight-powered structures $f$ is the
covering fraction of intercepted starlight; more generally it is simply the
dissipated power in stellar units. 

A grey radiator is one whose emissivity is independent of frequency, $\epsilon_\nu=\epsilon$ (equivalently $\beta=0$). More generally, thermal dust emission is often described by a frequency-dependent emissivity $\epsilon_\nu\propto\nu^\beta$, where $\beta$ is the dimensionless emissivity index. Natural cold dust typically exhibits $\beta\sim1$--2, while an ideal grey radiator has $\beta=0$.

The observed flux density follows from two simple steps. First, energy balance sets the size. By equation~(\ref{eq:area}), a spherical
shell rejecting $f L_{*}$ at temperature $T$ -- each unit area radiating
$\sigma T^{4}$ while absorbing $\sigma \Tcmb^{4}$ from the microwave
background in which it is immersed -- requires
$4\pi R^{2} = fL_{*}/[\epsilon\sigma(T^{4}-\Tcmb^{4})]$, i.e.\ radius
\begin{equation}
R(T,f)=\left[\frac{f L_{*}}{4\pi\epsilon\sigma\left(T^{4}-\Tcmb^{4}\right)}\right]^{1/2}.
\label{eq:radius}
\end{equation}

Second, brightness geometry sets the flux. A source of uniform specific
intensity $I_{\nu}$ subtending solid angle $\Omega = \pi R^{2}/d^{2}$ (its
projected disc) delivers $S_{\nu} = I_{\nu}\,\Omega$. Submillimetre and
millimetre measurements are differential: the sky glows everywhere with the
CMB, and an optically thick body also occults that background over its own
solid angle, so the measurable excess intensity is not
$\epsilon_{\nu}B_{\nu}(T)$ but
$\epsilon_{\nu}[B_{\nu}(T) - B_{\nu}(\Tcmb)] \equiv
\epsilon_{\nu}\,\Delta B_{\nu}(T)$ -- the excess-spectrum formalism
introduced by \citet[his equations 38--39]{lacki16} for near-CMB blackboxes.
Combining the radiator size from equation~(\ref{eq:radius}) with the brightness of a grey body gives the observed flux density \begin{equation} S_{\nu} = f\,\frac{\pi R^{2}(T,1)}{d^{2}}\, \epsilon_{\nu}\, \Delta B_{\nu}(T) = \frac{fL_{*}\,\epsilon_{\nu}\,\Delta B_{\nu}(T)} {4\,\epsilon\,\sigma \left(T^{4}-\Tcmb^{4}\right)d^{2}}. \label{eq:snu} \end{equation} 

The key result is that the flux depends only on the total dissipated power $fL_{*}$, the operating temperature $T$, and the distance $d$. It does not depend on how the emitting area is arranged. A complete shell, a sparse swarm, or any intermediate configuration with the same total power and temperature produces the same unresolved flux density. 

Note also that the emissivity at the observing frequency, $\epsilon_\nu$ need not equal the bolometric emissivity $\epsilon$ that enters the thermal balance. We therefore retain $\epsilon_\nu$ explicitly in the expression for the observed flux density. For most of this paper, however, we are interested in the simplest case of an optically thick grey radiator, for which the emissivity is approximately independent of wavelength. We therefore set $\epsilon_\nu=\epsilon=1$ throughout and for grey surfaces the
choice costs no generality: emissivity enters equation~(\ref{eq:snu}) only
through the ratio $\epsilon_{\nu}/\epsilon$, which is unity for any grey
radiator. A surface with $\epsilon = 0.5$ must deploy twice the area to
reject the same power, and so presents exactly the same flux. The assumption
that carries content is greyness itself -- $\epsilon_{\nu}$ approximately
constant across the band -- which is what distinguishes optically thick
engineered surfaces from optically thin natural dust, whose
modified-blackbody spectra have emissivity index $\beta \simeq 0.5$--2 (large
debris-disc grains at the low end, interstellar dust at the high end;
Section~\ref{sec:discriminants}); it
fails only for radiators engineered to be dark at precisely the observing
frequencies (Section~\ref{sec:caveats}). 

Finally, we note that equation~(\ref{eq:snu}) shows that, for a fixed temperature, the flux density depends only on the dissipated luminosity $fL_*$ and the distance $d$. The tabulated results presented in this paper therefore scale directly with $fL_*$ and as $d^{-2}$.

\begin{figure}
\centering
\includegraphics[width=\columnwidth]{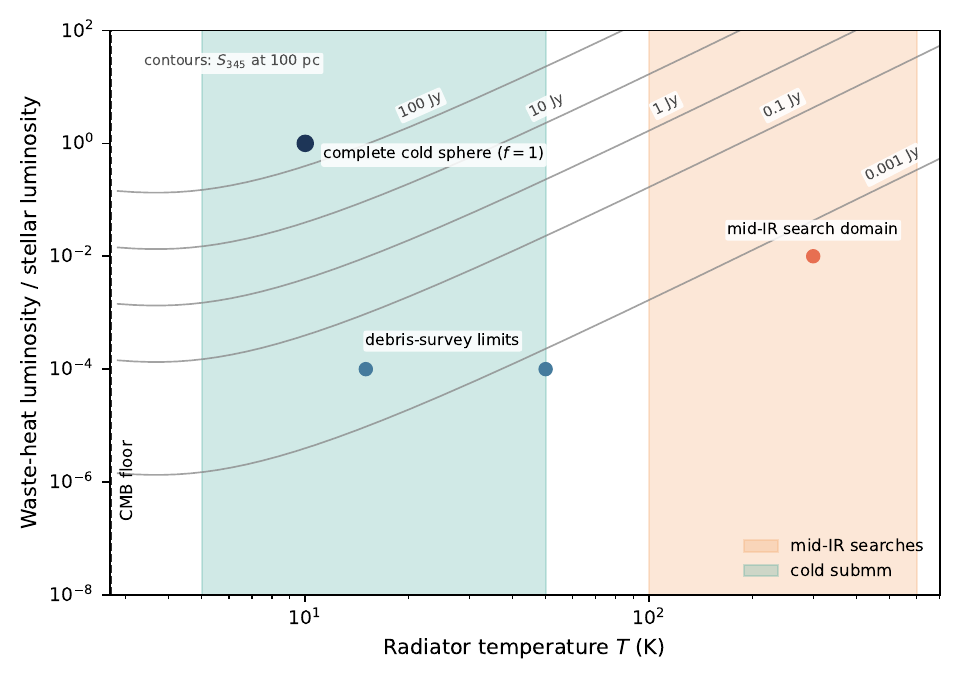}
\caption{Waste-heat temperature--luminosity phase space. Existing mid-infrared
searches occupy the warm band (orange); cold computational radiators fall in
the far-infrared/submillimetre band (green) towards the CMB floor. Grey lines
show 345-GHz flux-density contours for a source at 100~pc, from
equation~(\ref{eq:snu}), which is linear in the fractional waste-heat
luminosity $f$ on the vertical axis; the figure thus covers partial swarms
and complete spheres ($f=1$) alike. Markers indicate a complete $1\,\Lsun$ cold sphere,
the level of the archival debris-survey limits
(Section~\ref{sec:partial}), and the domain of warm mid-infrared searches.}
\label{fig:phase}
\end{figure}

Figure~\ref{fig:phase} locates the resulting search space: existing
mid-infrared programmes occupy the warm band, while Slysh haloes and
complete cold shells fall in the far-infrared/submillimetre band near the CMB
floor, at 345-GHz flux densities that are large by the standards of modern
surveys. Note that the figure extends into the regime $L_{\rm tech}/L_* > 1$. While starlight-powered haloes are confined to $L_{\rm tech}/L_* \leq 1$, the larger values shown here represent systems whose dissipation is supported by additional energy sources.

\section{Observability of Slysh haloes}
\label{sec:observability}

\subsection{Stellar and halo emission in different wavebands}

Stellar photospheres are intrinsically faint in the submillimetre. A solar-type star on the Rayleigh--Jeans tail presents only $\sim$0.3~mJy at 850~$\mu$m at a distance of 10~pc, while the nearby M dwarfs that dominate the local stellar population are typically an order of magnitude fainter still. For most systems relevant to this search, the photospheric contribution therefore lies well below the brightness of any detectable halo and is routinely modelled and subtracted in debris-disc studies. Only for the very nearest stars does the photosphere become directly measurable in its own right. At the distance of $\alpha$~Cen, a solar-type photosphere would appear as a $\sim$20~mJy source, and main-sequence photospheres are detected routinely across the ALMA bands \citep{liseau16}. Even in these cases, however, the stellar contribution is predictable, unresolved and spatially distinct from the extended halo emission considered here. A Slysh halo of
$\Ltech = 10^{21}$~W at 15~K around the same star is a $\sim$30~mJy source at
850~$\mu$m (Table~\ref{tab:fluxes}) - nearly two orders of magnitude
brighter than
its star at that wavelength, while contributing nothing detectable in the
optical. The system therefore presents complementary appearances in the optical and submillimetre: the star dominates at optical wavelengths, while the halo dominates in the submillimetre. 

The two components differ in shape as well as in colour. A photosphere is a
point source to any submillimetre facility -- $\alpha$~Cen~A, the largest
stellar disc in the sky after the Sun, subtends 8~mas -- whereas the halo is
extended on scales set by equation~(\ref{eq:teq}). A star observed alone
therefore typically returns an unresolved source of the predictable photospheric flux;
a star with a halo returns something brighter \emph{and broader than the
beam}, and the measured size is the second half of the signature. How large
the halo appears on the sky, and which facility is matched to it, depends
strongly on the host luminosity, and the next three subsections address this in more detail.

\subsection{Aggregate flux and detectability}

Table~\ref{tab:fluxes} gives the flux density of haloes of dissipation
$\Ltech = f\,\Lsun$ at ambient temperatures 10--30~K for stars at 10 and
100~pc, from equation~(\ref{eq:snu}); scaling is linear in $f$ and as
$d^{-2}$. Depths of $\sim$2~mJy (SCUBA-2 survey mode) correspond to
$\Ltech \sim 7\times10^{19}$~W at 10~pc and $\sim$$7\times10^{21}$~W at
100~pc.

\begin{table}
\centering
\caption{Slysh-halo flux densities from the grey-body model of
Section~\ref{sec:model} (equation~\ref{eq:snu}, $\epsilon = 1$;
equivalently a two-sided emitting area $f\Lsun/[2\sigma(T^{4}-\Tcmb^{4})]$
with mean projected cross-section half that).}
\label{tab:fluxes}
\begin{tabular}{lcccc}
\toprule
$f = \Ltech/\Lsun$ & $T$ (K) & $d$ (pc) & $S_{345\,\rm GHz}$ & $S_{857\,\rm GHz}$ \\
\midrule
$10^{-6}$ & 10 & 10  & 25 mJy   & 27 mJy \\
$10^{-6}$ & 12 & 10  & 17 mJy   & 27 mJy \\
$10^{-6}$ & 15 & 10  & 11 mJy   & 22 mJy \\
$10^{-6}$ & 30 & 10  & 1.8 mJy  & 6.9 mJy \\
$10^{-4}$ & 15 & 10  & 1.1 Jy   & 2.2 Jy \\
$10^{-4}$ & 15 & 100 & 11 mJy   & 22 mJy \\
$10^{-2}$ & 10 & 10  & 255 Jy   & 275 Jy \\
$10^{-2}$ & 15 & 100 & 1.1 Jy   & 2.2 Jy \\
$10^{-2}$ & 30 & 100 & 180 mJy  & 690 mJy \\
\bottomrule
\end{tabular}
\end{table}

\subsection{Halo angular scale} \label{sec:resolution} The flux densities of Table~\ref{tab:fluxes} are integrated quantities. For luminous hosts, however, that flux is distributed over a very large solid angle. The angular size of a halo follows directly from equation~(\ref{eq:teq}), with the characteristic radius scaling as $r\propto L_*^{1/2}$ and the angular extent as $\theta\propto L_*^{1/2}/d$. This produces a wide range of observable sizes. By a happy coincidence, a 12-K halo around a nearby M dwarf typically spans only $40''$--$4''$ over distances of 1--10~pc, corresponding to one to several JCMT beams at 850~$\mu$m. Such structures may be only marginally resolved, or even unresolved, in single-dish observations. By contrast, the 100--3000-au halo zone around a solar-luminosity star extends over $10''$--$300''$ at the same distance. Haloes around nearby G- and K-type stars therefore occupy scales of several arcminutes and are better matched to the beam sizes of archival all-sky surveys such as \textit{Planck} \citep{planck18hfi}. 

Once a halo is resolved, surface brightness becomes a more useful quantity than integrated flux density. Material at temperature $T$ occupies the radius $r(T)$ given by equation~(\ref{eq:teq}); combining equation~(\ref{eq:snu}) with the projected solid angle $\Omega=\pi r^2(T)/d^2$ gives \begin{equation} I_{\nu}\simeq4f\,\Delta B_\nu(T), \label{eq:sb} \end{equation} where $f$ is the covering fraction. The omitted factor $T^4/(T^4-T_{\rm CMB}^4)$ differs from unity by less than one per cent for $T \ga 9$~K, rising to $\sim$10 per cent at 5~K. The result is notable: for a halo of fixed temperature and covering fraction, the surface brightness is independent of both distance and host luminosity. Increasing the distance reduces the total flux density but decreases the angular area by the same factor, leaving the brightness per beam unchanged. This behaviour has two consequences. First, the detectability of a resolved halo is driven primarily by instrumental surface-brightness sensitivity rather than distance itself. Secondly, low-luminosity hosts enjoy an important observational advantage: because their haloes are intrinsically compact, a larger fraction of the emission remains concentrated within a small number of beams. The relationship between halo size, host luminosity and observing facility is explored further in the following subsections.

\subsection{M dwarfs as preferred Slysh halo targets}
\label{sec:hostL}

Equation~(\ref{eq:teq}) defines the halo zone by temperature rather than radius, so its geometry follows the luminosity of the host star. At fixed temperature the characteristic radius scales as $r\propto L_*^{1/2}$. Table~\ref{tab:hosts} illustrates the resulting trend across the main sequence. Around an M5 dwarf the full 5--28~K temperature range occupies a region extending from only a few au to about 120~au, whereas around a luminous A star the same thermodynamic regime lies hundreds to many thousands of au from the host. The CMB temperature floor and the onset of passive superconducting operation likewise occur much closer to low-luminosity stars. For niobium, for example, the transition temperature is reached at only $\sim35$~au around an M5 dwarf but near 900~au around the Sun.

\begin{table}
\centering
\caption{The halo zone as a function of host luminosity, from
equation~(\ref{eq:teq}). Radii scale as $L_{*}^{1/2}$ at fixed temperature;
$\theta_{12}$ is the angular radius of the 12-K case at a common distance of
5~pc, for comparison with beams of $7''$--$18''$ (\textit{Herschel}),
$14.6''$ (JCMT; \citealt{dempsey13}) and $4.7'$ (\textit{Planck} at 545~GHz).}
\label{tab:hosts}
\setlength{\tabcolsep}{4pt}
\begin{tabular}{lrrrrr}
\toprule
Type & $L_{*}/\Lsun$ & $r_{28}$ & $r_{12}$ & $r_{5}$ & $\theta_{12}$ \\
     &               & (au)     & (au)     & (au)    & (arcsec) \\
\midrule
A5V & 12      &  342 & 1863 & 10730 & 373 \\
F5V & 3.2     &  177 &  962 &  5540 & 192 \\
G2V & 1.0     &   99 &  538 &  3100 & 108 \\
K5V & 0.15    &   38 &  208 &  1200 &  42 \\
M0V & 0.07    &   26 &  142 &   820 &  28 \\
M3V & 0.010   &   10 &   54 &   310 &  11 \\
M5V & 0.0015  &    4 &   21 &   120 &   4 \\
\bottomrule
\end{tabular}
\end{table}

The compact geometry of dwarf-star haloes produces a useful observational advantage. At a given temperature the angular size scales as $\theta\propto L_*^{1/2}/d$, while the flux density depends only on the technological power, temperature and distance (equation~\ref{eq:snu}). Host luminosity does not enter explicitly. A halo around an M dwarf is therefore smaller on the sky than an equivalent structure around a solar-type or A-type star, but it is not intrinsically fainter. The surface brightness is also correspondingly higher and less susceptible to beam dilution. Resolved haloes around luminous nearby stars may extend over many beam widths, whereas comparable structures around M dwarfs can remain compact even for the nearest systems. This observational advantage is reinforced by stellar demographics. M dwarfs constitute roughly three quarters of the stellar population within 10~pc and dominate the solar neighbourhood. Nearby examples such as Proxima Centauri, Barnard's Star and Wolf~359 therefore provide some of the closest potential halo targets. At the same time, the compact angular scales predicted in Table~\ref{tab:hosts} are well matched to the resolution of existing submillimetre survey data. While \textit{Planck} provides an all-sky archival resource, higher-resolution facilities such as JCMT and \textit{Herschel} are particularly well suited to identifying compact cold haloes around nearby low-mass stars. 

There are also reasons to suspect that long-lived dwarf stars may be preferred hosts from the perspective of the civilisation itself. In an energy-limited picture, the total amount of computation available scales with the integrated stellar output $\int L_*\mathrm{d}t$. Since the main-sequence lifetime varies approximately as $M_*/L_*$, this quantity changes much less dramatically across the main sequence than luminosity alone. A mid-M dwarf may provide a substantial fraction of the Sun's lifetime energy budget, but spread over $\sim10^{12}$ rather than $\sim10^{10}$~yr. Over such timescales the cosmic microwave background continues to cool, progressively reducing the Landauer cost of computation. If mature technological systems are ultimately constrained by available energy rather than processing rate, long-lived low-mass stars may be among the most attractive hosts. They are also the stars around which the submillimetre signature of a Slysh halo may be easiest to detect.


\subsection{Matching halo scale to observing facility} 
\label{sec:beams} 
The principal observational challenge is not sensitivity alone but the relationship between halo size and telescope beam. Equation~(\ref{eq:teq}) shows that the characteristic angular scale depends on both host luminosity and distance, while equation~(\ref{eq:snu}) determines the integrated flux density. Different facilities therefore occupy different regions of the parameter space. Nearby M-dwarf systems are particularly interesting. Their 12-K haloes typically span only a few tens of arcseconds (Table~\ref{tab:sizes}), corresponding to one to several JCMT beams at 850~$\mu$m. These structures are large enough to reveal extension beyond the stellar photosphere but compact enough to avoid severe dilution over many beams. For $f=10^{-5}$ the predicted halo flux densities exceed the stellar photospheric contribution by one to two orders of magnitude. Proxima Centauri is especially attractive: a 2-mJy sensitivity limit already probes technological powers of order $\Ltech \sim 7\times10^{17}$~W. Solar-type and more luminous stars occupy a different regime. Their halo zones extend to hundreds or thousands of au, producing angular scales of several arcminutes for the nearest systems. Such haloes are increasingly resolved by single-dish facilities and are very poorly matched to the compact structures targeted by interferometers. Archival all-sky surveys remain valuable in this regime. The nearest G- and K-type stars are predicted to host haloes with angular sizes comparable to the beam of \textit{Planck}, making existing survey data a natural resource for searches around the brightest nearby hosts. 

Interferometers provide the complementary capability. As distance increases, the angular size of a halo falls as $d^{-1}$ while its total flux density decreases as $d^{-2}$. Structures spanning tens of arcseconds around the nearest stars shrink to only a few arcseconds at distances of tens to hundreds of parsecs. In this regime facilities such as ALMA and NOEMA become increasingly well matched to the expected source size. Their combination of angular resolution and continuum sensitivity makes them natural instruments for both the discovery and characterisation of compact haloes around low-luminosity stars. Indeed, the observational advantages identified in Section~\ref{sec:hostL} extend across a wide range of distances. Haloes around M dwarfs are intrinsically compact, yet their flux density depends only on technological power, temperature and distance rather than host luminosity. Consequently, low-mass stars remain attractive targets even when they are no longer among the nearest systems. The same compactness that makes nearby M-dwarf haloes suitable for JCMT observations makes more distant examples well matched to ALMA and NOEMA. The practical observing strategy is therefore hierarchical. Existing survey data can be used to identify large-scale emission around nearby luminous stars, while nearby M dwarfs are natural targets for sensitive single-dish observations. Interferometers such as ALMA and NOEMA extend the search volume to much larger distances and provide the means to resolve candidate haloes, distinguishing annular structures, warm inner nodes and departures from azimuthal symmetry that would appear only as excess size or surface brightness in lower-resolution data (Section~\ref{sec:discriminants}).

\begin{table}
\centering
\caption{The 12-K case around the nearest stars
(equation~\ref{eq:teq}). Diameters are $2r_{12}/d$;
$S_{*}$ is the 850-$\mu$m photospheric flux density (blackbody at
$T_{\rm eff}$; true submillimetre brightness temperatures are somewhat
lower) and $S_{\rm halo}$ the total halo flux density at $f = 10^{-5}$,
$T = 12$~K (equation~\ref{eq:snu}). M dwarf haloes are matched to the
$14.6''$ JCMT beam, K and G haloes to the $\sim$$5'$ \textit{Planck}
beams.}
\label{tab:sizes}
\setlength{\tabcolsep}{3pt}
\begin{tabular}{llrrrrr}
\toprule
Star & SpT & $d$ & $L_{*}/\Lsun$ & $r_{12}$ & Diam. & $S_{*}$ / $S_{\rm halo}$ \\
     &     & (pc) &               & (au)     &                & (mJy) \\
\midrule
Proxima Cen    & M5.5V & 1.30 & 0.0016 & 21  & $33''$  & 0.3 / 16 \\
Barnard's star & M4V   & 1.83 & 0.0035 & 32  & $35''$  & 0.2 / 19 \\
Wolf 359       & M6V   & 2.41 & 0.0014 & 20  & $17''$  & 0.1 / 4.3 \\
Lalande 21185  & M2V   & 2.55 & 0.023  & 82  & $64''$  & 0.5 / 63 \\
Ross 154       & M3.5V & 2.98 & 0.0038 & 33  & $22''$  & 0.1 / 7.6 \\
$\alpha$ Cen A & G2V   & 1.34 & 1.52   & 663 & $16.5'$ & 29 / 15000 \\
61 Cyg A       & K5V   & 3.50 & 0.15   & 210 & $2.0'$  & 1.0 / 220 \\
$\epsilon$ Eri & K2V   & 3.22 & 0.34   & 314 & $3.2'$  & 1.6 / 580 \\
$\tau$ Cet     & G8V   & 3.65 & 0.52   & 388 & $3.5'$  & 1.5 / 690 \\
\bottomrule
\end{tabular}
\end{table}

\subsection{Radial structure and the temperature law}
\label{sec:structure}

If a halo operates at the local ambient temperature, its thermal
emission must follow the equilibrium relation of
equation~(\ref{eq:teq}), with material becoming progressively colder at
larger radii. The observational consequence is a radial colour
gradient: the inner halo appears warmer than the outer halo, while each
annulus radiates approximately as a grey body. A multi-band image
therefore provides more than a simple detection. It tests two specific
predictions simultaneously: nearly grey emission ($\beta \approx 0$)
and the temperature law $T \propto r^{-1/2}$. Known circumstellar dust
populations do not naturally satisfy both conditions. Debris discs
follow a broadly similar temperature law, but their millimetre
emissivity indices are typically $\beta \approx 0.5$--1
\citep{macgregor16} and they are commonly detected in scattered light.

The radial brightness profile contains further information about the
distribution of infrastructure. In a halo where comparable amounts of
power are dissipated at all radii, the emission is spread relatively evenly across the full halo
zone, spanning the range of radii over which the ambient temperature
falls from roughly 30 K to a few kelvin. For a solar-luminosity
star this corresponds to approximately 100--3000 au, but the scale
contracts as $L_*^{1/2}$ for lower-luminosity hosts.
Concentrating the same technological
power preferentially toward the inner or outer halo instead produces a
much more centrally concentrated or extended appearance. This differs
from a conventional debris belt, whose emission is usually confined to a
much narrower range of radii and therefore exhibits a correspondingly
peaked radial profile. Spatial structure therefore provides an
additional discriminant (see Section~\ref{sec:discriminants}) alongside spectral slope and temperature.

The apparent size of a halo depends on observing wavelength. Shorter wavelengths preferentially sample the warmer inner regions, while longer wavelengths become increasingly sensitive to the colder material at larger radii. Consequently a halo generally appears more compact in the far-infrared and progressively more extended towards millimetre wavelengths. Multi-wavelength imaging therefore provides a direct probe of both the temperature structure and the radial distribution of the emitting material, while also determining what fraction of the total emission is recovered by a particular observing facility (Section~\ref{sec:limits}).

\section{Limits from archival data} 
\label{sec:limits} 

\subsection{Partial swarms: debris-disc surveys as Dysonian datasets} 
\label{sec:partial}

The observational signatures derived in Section \ref{sec:observability} lie squarely within the parameter space explored by existing far-infrared and submillimetre surveys, making their archives a natural starting point for an initial search for Slysh haloes. Unlike classical mid-infrared Dyson-sphere searches, which target waste heat near habitable-zone temperatures, the relevant temperature range here is approximately 5--30 K, placing the peak emission between the far-infrared and millimetre bands. A wide range of surveys have already explored this parameter space for other scientific purposes, including debris-disc programmes around nearby stars and all-sky submillimetre surveys. Figure~\ref{fig:phase} places representative archival observations on the temperature--luminosity plane introduced in Section~4. 

Existing facilities probe a substantial fraction of the Slysh-halo parameter space. The deepest observations of nearby stars reach sensitivities that, in principle, correspond to technological powers of order $10^{20}$~W, while bright cold structures with covering fractions approaching unity would be detectable over much larger volumes. These numbers should be regarded as indicative rather than definitive - the observations were not designed as technosignature surveys, and the translation from survey sensitivity to halo constraints depends on details of source extraction, angular filtering, photospheric subtraction, confusion and catalogue construction. For nearby stars, archival debris-disc programmes are particularly relevant. Surveys such as \textit{Herschel} DEBRIS and DUNES and the JCMT SONS programme (see Table \ref{tab:surveys}) targeted hundreds of stars within a few tens of parsecs and achieved milliJansky sensitivities in the wavelength range where cold haloes are expected to radiate. In the absence of a dedicated re-analysis, the prudent conclusion is not that these surveys exclude Slysh haloes at a specific level, but that they already contain data capable of testing a not insignificant region of parameter space.

\begin{figure}
\centering
\includegraphics[width=\columnwidth]{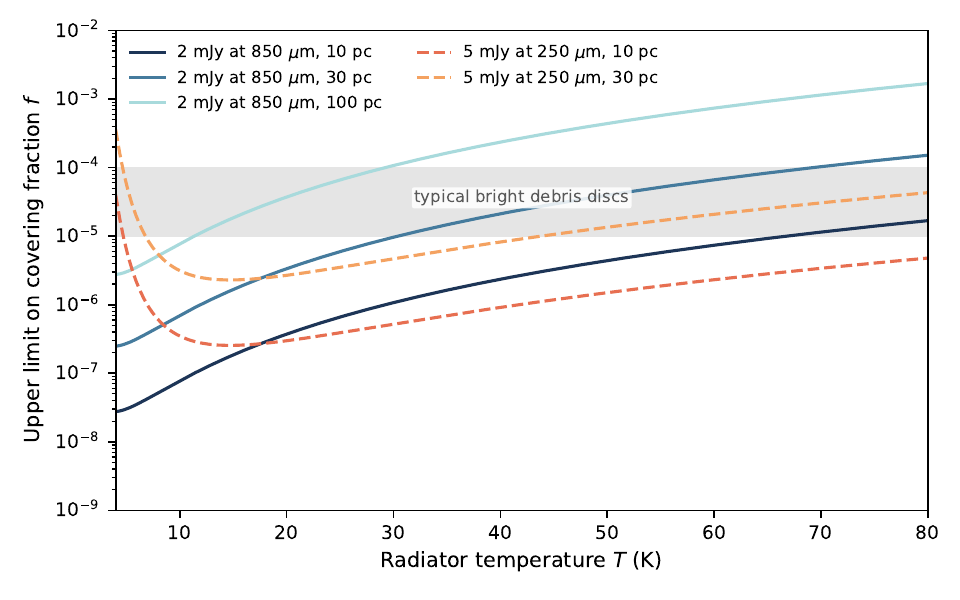}
\caption{Illustrative covering-fraction limits for partial cold swarms from
non-detections at 850~$\mu$m (2~mJy) and 250~$\mu$m (5~mJy) around a
$1\,\Lsun$ star at 10, 30 and 100~pc (equation~\ref{eq:flim}, grey radiators).
The shaded band marks the fractional luminosities of typical bright debris
discs. These model curves should be replaced by per-star limits using the
actual noise, distance, luminosity and bandpass of each observation.}
\label{fig:flimits}
\end{figure}

The form such a re-analysis would take is simple. Inverting equation~(\ref{eq:snu}), a non-detection at limiting flux density $S_{\nu,\rm lim}$ translates into an upper limit on the covering fraction of grey radiators at temperature $T$,
\begin{equation}
f_{\rm lim}(T)= \frac{S_{\nu,\rm lim}\,d^{2}\,
4\sigma\left(T^{4}-\Tcmb^{4}\right)}{L_{*}\,\Delta B_{\nu}(T)},
\label{eq:flim}
\end{equation}
evaluated star by star with the actual noise, distance and luminosity of each observation. Figure~\ref{fig:flimits} illustrates the limits this yields at representative survey depths: fractional dissipations of $10^{-7}$--$10^{-5}$ are within reach around 10-pc stars, two to three orders of magnitude below the fractional luminosities of even bright debris discs. A complementary regime is provided by shallow all-sky surveys and deep observations of specific fields.

\begin{table}
\centering
\caption{Archival debris-disc surveys reinterpreted in this work
\citep{matthews10,eiroa13,holland17}. Sensitivities are representative
1$\sigma$ point-source values; per-target depths vary. Sensitivities refer to the primary band: 100~$\mu$m for DEBRIS/DUNES (160-$\mu$m depths are $\sim$2--3$\times$ shallower), 850~$\mu$m for SONS.}
\label{tab:surveys}
\setlength{\tabcolsep}{4pt}
\begin{tabular}{lcccc}
\toprule
Survey & $N_{*}$ & $d$ & $\lambda$ & $\sigma_{\nu}$ \\
       &         & (pc) & ($\mu$m) & (mJy) \\
\midrule
DEBRIS$^{a}$ & 446 & $\la$45 & 100/160 & $\sim$1.2 \\
DUNES$^{b}$  & 133 & $\la$25 & 100/160 & $\sim 1.5$ \\
SONS$^{c}$   & 100 & $\la$50 & 850 (450) & 1.4 \\
\bottomrule
\end{tabular}
\begin{minipage}{\columnwidth}
\vspace{4pt}
\footnotesize
$^{a}$Unbiased flux-limited census of the nearest A--M stars, observed to
uniform depth. 
$^{b}$Deepest photosphere-limited search for Solar-System-like cold belts
around the nearest Sun-like (FGK) stars.
$^{c}$Submillimetre follow-up, preferentially targeting known and suspected
disc hosts; the strongest constraints on the coldest material.
\end{minipage}
\end{table}

\subsection{The complete-shell limiting case}
\label{sec:shell}

\begin{figure}
\centering
\includegraphics[width=\columnwidth]{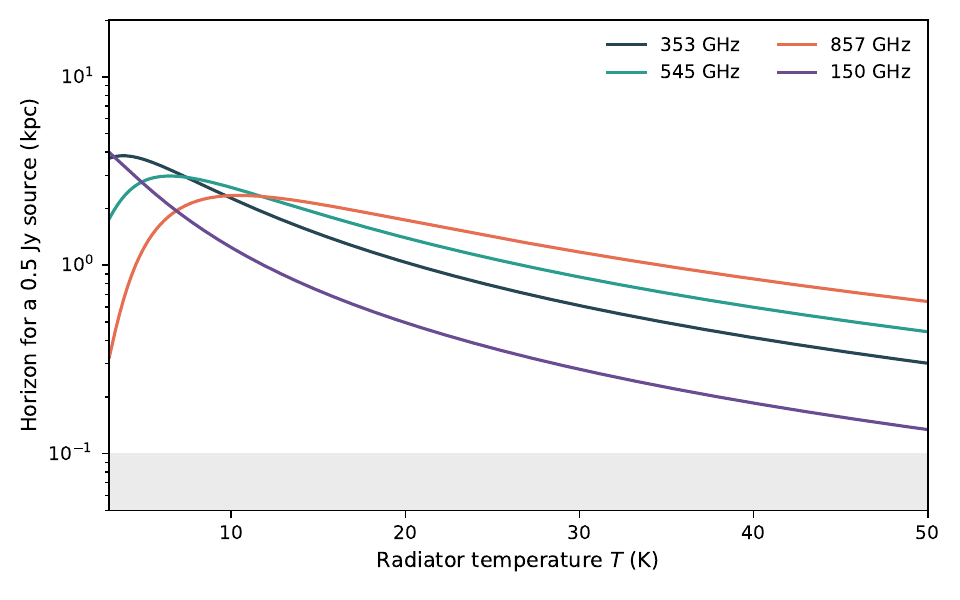}
\caption{Detection horizon for a complete $1\,\Lsun$ cold sphere as a function
of radiator temperature, for a 0.5-Jy flux-density limit at four
representative bands (equation~\ref{eq:horizon}). The broad maximum moves with
observing band; the key point is that horizons are kiloparsec-scale over much
of the 5--20~K range.}
\label{fig:horizon}
\end{figure}

Complete cold shells ($f\rightarrow1$) are remarkably conspicuous.
Table~\ref{tab:complete} evaluates equations~(\ref{eq:radius}) and
(\ref{eq:snu}) for $1\,\Lsun$ structures at 5--30~K, at 100~pc and 1~kpc, in
four representative millimetre and submillimetre bands. Three features stand
out. The flux densities are large even out to 100~pc, and the brightest band remains at or near the Jy
level even at 1~kpc. The band of peak flux marches with temperature - the maximum flux moves
from 353~GHz for a 5-K shell, through 545~GHz at 8--10~K, to 857~GHz by
15--30~K. The same effect produces the crossing of the horizon curves in
Fig.~\ref{fig:horizon}. And since
$R \propto T^{-2}$ (equation~\ref{eq:radius}), the coldest shells are also
the largest. 

For solar-luminosity hosts their expected emission falls naturally into the \textit{Planck} frequency range, and such objects would be expected to enter compact source catalogues over large distances. Whether existing catalogued cold sources already exclude this possibility is uncertain in our view -  classification rather than raw sensitivity is the limiting issue. All but the very nearest sources are unresolved by \textit{Planck},
whose 545-GHz beam is $\sim$4.7~arcmin. At 1~kpc complete shells remain a several-Jy source, and at distances beyond a few hundred parsecs they are unresolved by ground-based single dishes. Interferometers such as ALMA and NOEMA would begin to resolve these sources, and could establish their spatial morphology (see also section \ref{sec:discriminants}).   

For a compact-source threshold $S_{\rm lim}$, the maximum distance at which a shell can be detected follows directly from the inverse-square law: 
\begin{equation}
d_{\rm hor}=100\,{\rm pc}
\left[\frac{S_{\nu}(100\,{\rm pc})}{S_{\rm lim}}\right]^{1/2},
\label{eq:horizon}
\end{equation}
which for a representative $S_{\rm lim}=0.5$~Jy at 545~GHz gives
$d_{\rm hor} \simeq 2.6\,(L_{*}/\Lsun)^{1/2}$~kpc at $T=10$~K and kiloparsec-scale horizons
across the whole 5--20~K range (Fig.~\ref{fig:horizon}). The $L_{*}^{1/2}$
scaling carries the dwarfs with it: a complete shell around an M5 dwarf,
which by Section~\ref{sec:mass} costs only a fifth of an Earth mass, would
be catalogued out to $\sim$100~pc. 

The band dependence in Fig.~\ref{fig:horizon} is itself informative: each
frequency reaches farthest for shells whose spectra peak within it, so the
lowest frequencies carry the search closest to the CMB floor (150 and
353~GHz dominate below $\sim$7~K) while 857~GHz takes over for warmer
shells. The ratio of horizons between bands is another expression of the colour
information exploited by the spectral-greyness test of
Section~\ref{sec:discriminants}.

The representative 0.5-Jy threshold adopted above is, in fact, close to the
published 90 per cent completeness levels of the Second \textit{Planck}
Catalogue of Compact Sources in the extragalactic zone
\citep[PCCS2; 555~mJy at 545~GHz, 791~mJy at 857~GHz;][]{planck16pccs2} -- for which the 10-K horizon
becomes 2.5~kpc (1.9~kpc at 857~GHz). The exact reach varies with sky
position, beam dilution, source extraction and calibration, but the
conclusion is robust: solar-luminosity complete cold spheres are not subtle.
 If they exist within kiloparsec distances, they
are already present in all-sky submillimetre catalogues -- present, but possibly not
yet identified. 

Distinguishing a true Slysh halo from molecular clouds, Galactic cold clumps and other natural populations requires the discriminants developed in the next section. The principal conclusion of this section is therefore not that existing surveys have already placed robust limits on Slysh haloes, but that the necessary observations largely exist, at least for complete shells. Archival far-infrared and submillimetre datasets already reach levels that are astrophysically interesting for nearby stars and potentially sensitive to complete cold shells over very large Galactic volumes. The relevant question is not whether the PGCC contains cold objects - it does - but whether any object has the wrong \emph{kind} of coldness. 
Establishing quantitative constraints requires a dedicated analysis designed specifically for the halo hypothesis rather than the scientific objectives for which the data were originally obtained.

\begin{table*}
\centering
\caption{Complete grey Dyson spheres ($f=1$) with $L_{\rm waste}=1\,\Lsun$,
from equations~(\ref{eq:radius}) and (\ref{eq:snu}). The CMB enters both the
luminosity balance and the observed contrast $\Delta B_{\nu}$. The angular
radius is $R/d$ in arcsec, since 1~AU at 1~pc subtends 1~arcsec. Columns are
ordered by increasing frequency, so the Wien march of the row maxima is
visible directly: the brightest band steps from 353~GHz at 5~K to 857~GHz by
15--30~K.}
\label{tab:complete}
\begin{tabular}{rrrrrrrr}
\toprule
$T$ & $R$ & $d$ & $\theta$ & $S_{150}$ & $S_{353}$ & $S_{545}$ & $S_{857}$ \\
(K) & (au) & (pc) & (arcsec) & (Jy) & (Jy) & (Jy) & (Jy) \\
\midrule
5  & 6491 & 100  & 64.9 & 362.0 & 664.7 & 394.1 & 77.2 \\
5  & 6491 & 1000 & 6.49 & 3.62  & 6.65  & 3.94  & 0.77 \\
8  & 2437 & 100  & 24.4 & 132.8 & 383.4 & 413.2 & 239.6 \\
8  & 2437 & 1000 & 2.44 & 1.33  & 3.83  & 4.13  & 2.40 \\
10 & 1554 & 100  & 15.5 & 77.3  & 257.9 & 335.4 & 275.1 \\
10 & 1554 & 1000 & 1.55 & 0.77  & 2.58  & 3.35  & 2.75 \\
15 & 689  & 100  & 6.89 & 27.0  & 108.1 & 177.3 & 224.1 \\
15 & 689  & 1000 & 0.69 & 0.27  & 1.08  & 1.77  & 2.24 \\
30 & 172  & 100  & 1.72 & 3.93  & 18.7  & 37.5  & 69.1 \\
30 & 172  & 1000 & 0.17 & 0.04  & 0.19  & 0.38  & 0.69 \\
\bottomrule
\end{tabular}
\end{table*}

\section{Discriminants and the confusion budget}
\label{sec:discriminants}

The observational challenge posed by Slysh haloes is not simply one of detection but of discrimination. Existing far-infrared and submillimetre archives provide valuable opportunities to search for both partial haloes around nearby stars and complete shells over large Galactic volumes, but any candidate must ultimately be distinguished from the rich population of natural cold sources that populate the submillimetre sky. At the same time, the compact haloes expected around nearby and moderately distant M dwarfs are natural targets for new observations with facilities such as JCMT, ALMA and NOEMA, where improved sensitivity and angular resolution can provide decisive tests. The criteria required to interrogate archival catalogues are therefore largely the same as those needed to identify and confirm candidates in targeted observations. We outline six observational discriminants:

\textit{(i) Spectral greyness.} Engineered radiators are optically thick grey
bodies {\it i.e.} the model assumption of Section~\ref{sec:model} with
$\beta \approx 0$ and Rayleigh--Jeans slope $S_\nu \propto \nu^{2}$.
Optically thin cold dust is much steeper: interstellar dust and prestellar cores
show $\beta \simeq 1.5$--2 ($S_\nu \propto \nu^{3.5-4}$), and even the large
grains of debris discs retain $\beta \approx 0.5$--1
\citep{macgregor16,hughes18}, measurably distinct from zero. With
$h\nu/k = 16.9$, 26.2 and 41.1~K, the \textit{Planck} 353/545/857-GHz bands
straddle the spectral peak of a 5--20~K source, so the measurement is a joint
fit of $T$ and $\beta$ to $S_{\nu} \propto \nu^{\beta}\Delta B_{\nu}(T)$
through the real bandpasses; the three bands fix $T$ well and leave a
$T$--$\beta$ covariance that longer wavelengths break. Follow-up observations at 450/850~$\mu$m (JCMT/SCUBA-2) together with submillimetre and millimetre photometry from ALMA or NOEMA make the spectral slope decisive.

\textit{(ii) The radial law.} $T(r) \propto r^{-1/2}$ across a resolved halo
(equation~\ref{eq:teq}; observational form in
Section~\ref{sec:structure}) -- a two-parameter physical fit that no
known natural population is expected to pass in combination with (i). The
surface-brightness profile is a second handle on the same map: a halo spreads
its flux across a wide range of radii where a debris belt concentrates it near
one.

\textit{(iii) Line-free continuum.} Cold natural sources at 6--20~K are
embedded in molecular gas and bright in CO and dense-gas tracers
\citep{bergin07}; one pointing per candidate should eliminate most natural cores. A
line-free continuum source at 10~K is unusual.

\textit{(iv) Environmental and counterpart tests.} Prestellar cores and cirrus
knots reside in hierarchical cloud structure, correlate with H\,{\sc i} and
dust column-density maps, and cluster toward the Galactic plane and known
star-forming regions \citep{bergin07,planck16pgcc}; a shell should be
isolated, compact and often at high
latitude. A complete shell should also lack any \textit{Gaia}, 2MASS or WISE
counterpart at its centre, the 22-$\mu$m Wien tail
being suppressed relative to the 545-GHz flux density by factors of
$\sim$$10^{5}$ at 30~K. A halo should
lack the scattered light that accompanies dust (radiators need not reflect).

\textit{(v) Interferometric morphology.} Interferometers such as ALMA and NOEMA are particularly well suited to the compact haloes expected around nearby and moderately distant M dwarfs. The combination of high sensitivity and for ALMA configurable angular resolution, allows candidate sources to be traced across a wide range of spatial scales, from unresolved compact emission to resolved halo structure. A genuine halo is expected to exhibit thermal morphology on scales set by the host luminosity and the equilibrium temperature law of equation~(\ref{eq:teq}), while background AGN and many dusty galaxies typically remain dominated by a compact central component. The spatial structure of Slysh halo candidates becomes a diagnostic in its own right - compact candidates can be followed from arcsecond scales down to tens of milliarcseconds, while more extended systems can be tested for annular structure, radial gradients and departures from azimuthal symmetry. Interferometric observations therefore provide a powerful means of distinguishing compact thermal haloes from unrelated background sources, while simultaneously probing the internal structure of any detected emission and its spectra.

\textit{(vi) Co-motion.} For haloes the test is positive: the cold emission
must share the host star's \textit{Gaia} parallax and proper motion. This is
the astrometric discriminant of \citet{garrett26b} transplanted to the thermal
domain, and it is decisive against the dominant point-like contaminant --
high-redshift dusty galaxies, which are fixed on the sky. 
For complete shells (no star) the same test applies to the cold source
itself.
A tangential motion of 5~km\,s$^{-1}$ at 100~pc is only 10~mas\,yr$^{-1}$ - beyond single-dish
survey astrometry but possibly within reach of ALMA, whose phase-referenced
astrometry attains milliarcseconds per epoch: in principle, a long multi-epoch campaign could confirm or refute a candidate within a decade or two. 

The false-positive problem in waste-heat searches should not be underestimated. Recent work on the Hephaistos candidates, for example, has shown that at least some promising waste-heat signatures can be explained by contamination from background dust-obscured galaxies \citep{ren25}. Similar issues arise throughout the submillimetre sky, which contains a rich variety of natural cold sources spanning debris discs, prestellar cores, molecular clouds, young stellar objects, Galactic cirrus and high-redshift dusty galaxies. The purpose of the discriminant suite outlined above is therefore not merely to confirm candidates but to eliminate these often numerous and convincing false positives. The confusion budget for deep surveys is, in approximate order of importance, as follows. High-redshift dusty star-forming galaxies and AGN are likely to be the dominant point-like contaminants, particularly in blind surveys. They are fixed on the sky and often retain compact structure on long interferometric baselines, making discriminants (v) and (vi) especially powerful. Cold debris discs share some aspects of the halo temperature regime but occupy a different radial zone ($\la100$~au), exhibit $\beta\approx0.5$--1 rather than $\beta\approx0$, and scatter starlight: discriminants (i), (ii) and (iv). Resolved discs also commonly show rings, gaps, asymmetries and stellocentric offsets \citep{hughes18}, whereas a Slysh halo traces the distribution of infrastructure and need not resemble a collisional dust population. Prestellar cores, molecular clouds and Galactic cirrus are generally identified through their steeper spectral slopes, rich molecular-line spectra and cloud-associated environments: discriminants (i), (iii) and (iv). Evolved stars and young stellar objects can also produce cold thermal emission but are usually accompanied by bright infrared counterparts and broader spectral energy distributions. The discriminant suite is therefore not an appendix to the search programme; it is the search programme. No single test is likely to be decisive in every case, but the combination of spectral slope, temperature structure, environmental context, morphology, line properties and astrometric behaviour provides a set of largely independent filters that very few natural sources are expected to pass simultaneously.


\section{A three-tier Slysh halo search programme}
\label{sec:programme}

The searches above organise into three tiers of increasing cost; candidates
from any tier are classified with the discriminants of
Section~\ref{sec:discriminants} applied cumulatively. 

The first tier is archival and requires no new telescope time. It begins with a re-analysis of the DEBRIS, DUNES, SONS and related surveys, translating their photometry into star-by-star constraints in $(f,T,\epsilon)$ parameter space. It also includes an automated screen of the \textit{Planck} cold-clump and compact-source catalogues against discriminants (i), (iv) and (vi), extending the blackbox search of \citet{lacki16} to the circumstellar regime. Additional opportunities are provided by archival millimetre and submillimetre observations from JCMT, ACT \citep{naess20} and SPT \citep{carlstrom11}, which can be searched for cold circumstellar emission associated with Galactic stars. Both the debris-survey re-analysis and the \textit{Planck} screen will be presented in forthcoming papers (Garrett et al., in preparation).

The second tier is a dedicated submillimetre observing programme. It begins with the nearest stellar systems, where the expected halo zones are best matched to various single-dish facilities (see Section~\ref{sec:beams}). 
Around Proxima, Barnard's Star, Wolf~359, Ross~154 and Lalande~21185, the 12-K halo zone subtends only a few JCMT beams, making integrated measurements feasible. For these low-luminosity hosts, sensitivities approaching $\Ltech\sim10^{18}$~W ($f\sim10^{-6}$) are within reach in modest integration times making these systems natural first targets for a dedicated survey. 
As distance increases, the apparent halo size decreases and the preferred instrument changes. Single-dish telescopes are best suited to large, extended haloes because they recover low surface-brightness emission on large angular scales, while interferometers such as ALMA and NOEMA become increasingly useful for compact, distant, or clumpy systems. The reprocessing of ALMA archival data and new observations across multiple wavelengths then provide the greyness, radial-profile and colour tests of Section~\ref{sec:discriminants}. Candidates emerging from either tier then proceed to standard follow-up through molecular-line spectroscopy, continuum colours and resolved imaging.

The third tier looks forward to future survey facilities. Wide-field millimetre and submillimetre surveys from the Fred Young Submillimeter Telescope (FYST) and the Simons Observatory \citep{simonsobs19} will extend existing searches to much larger Galactic samples and improve sensitivity to cold circumstellar emission. The proposed PRobe far-Infrared Mission for Astrophysics (PRIMA) would restore sensitive coverage of the far-infrared regime, providing access to halo temperatures of approximately 20--60~K that have remained largely unexplored since \textit{Herschel}. Complementary candidate lists may also emerge from the optical domain. \textit{Gaia}-underluminous stars \citep{zackrisson18} and vanishing-star candidates from VASCO (Vanishing and Appearing Sources during a Century of Observations) project \citep{villarroel20} identify systems with anomalous optical properties that merit scrutiny at much longer wavelengths. Objects that combine unusual optical dimming with a cold far-infrared or submillimetre excess would represent especially compelling targets for follow-up.

Across all three tiers, candidates can be tested against the discriminants of Section~\ref{sec:discriminants} and either rejected or promoted for further study. This approach delivers either the first credible cold technosignature candidates or the strongest observational limits yet placed on computation-dominated technological activity in the local Universe (Section~\ref{sec:null}).

\section{Discussion}
\label{sec:discussion}

\subsection{Relation to previous work}

Earlier research in this area has been discussed in Sections~\ref{sec:intro} and \ref{sec:thermo}: cold shells \citep{slysh85,timofeev00}, near-CMB blackboxes \citep{lacki16}, the migration and aestivation arguments \citep{cirkovic06,sandberg16}, and the thermodynamics of radiation applied to Dyson spheres \citep{wright20,wright23}. The observational focus of this work is therefore not the classical Dyson sphere but the more general expectation that thermodynamically optimised computation produces extended, cold waste-heat structures, even when only a small fraction of the host star's luminosity is harvested. The contribution of this paper is to identify and characterise the observational regime between the warm circumstellar searches of IRAS \citep{carrigan09} and WISE \citep{griffith15},  and the galaxy-scale blackbox searches of \citet{lacki16}. 

Cold circumstellar structures are associated with individual stellar systems, permitting direct comparison with stellar properties and astrophysical environment, while remaining bright enough to be detectable in the far-infrared and submillimetre. In contrast, galaxy-scale searches probe enormous volumes but must contend with the complexity of entire galaxies, while warm circumstellar searches focus on a temperature range that may correspond to relatively immature or rate-optimised technologies. Cold circumstellar searches instead probe the thermodynamic endpoint suggested by migration and aestivation arguments. Around the low-luminosity M dwarfs that dominate the local stellar census, these structures are also expected to lie on angular scales accessible to both modern single-dish telescopes and interferometers (Section~\ref{sec:beams}). Viewed in this way, the traditional Dysonian SETI question is shifted. Rather than asking what fraction of a civilisation's energy supply is captured, we ask how efficiently that energy is ultimately used. The characteristic waste temperature then determines both the observational wavelength and the spatial scale on which the technosignature is expected to appear. 

The contrasting conclusions of \citet{wright23} and the present work (Section~\ref{sec:objections}) can likewise be viewed as complementary rather than contradictory. If construction mass is the dominant constraint, waste heat remains comparatively warm and infrared searches are favoured. If the dominant constraint is the total energy available over cosmic timescales, waste heat migrates toward much lower temperatures and submillimetre searches become increasingly important. Together, these approaches bracket the temperature axis and provide observational coverage across a broad range of technological optimisation strategies.

\subsection{The evolutionary reading and the null result} 
\label{sec:null} 

For the migration proposal, waste-heat temperature functions as an evolutionary clock. Technological infrastructure is born warm, close to its host star and operating at planetary temperatures of a few hundred kelvin, before expanding outward toward progressively colder environments. Humanity may already be entering the first stages of this transition through the emergence of orbital data centres and space-based computing infrastructure \citep{suncatcher25,marcy26}. In this picture, mid-infrared searches probe an earlier phase of development, whereas submillimetre searches probe the mature endpoint, where thermodynamically optimised computation and long-lived technological populations may accumulate. The distinction may be amplified by timescale. At AI-era growth rates, the transition from planetary- to stellar-scale infrastructure could occur on timescales of centuries rather than millennia \citep{nachtrieb26,garrett26a}. Warm, compact waste heat may therefore be a relatively transient phenomenon compared with the gigayears over which cold circumstellar structures can persist. If so, the cold sky is not merely an additional search channel but a natural place to look for the oldest and most enduring technological systems. Completion of the three-tier programme described in Section~\ref{sec:programme} would place substantially stronger observational constraints on this possibility. In the null case, it would establish stringent limits on complete cold spheres (5--30~K, $L\geq\Lsun$) across much of the nearby Galactic volume, while showing that the nearest stellar systems host no halo above $f\sim10^{-6}$--$10^{-4}$ over substantial temperature ranges. The archival tier alone would provide star-by-star constraints on halo luminosity fraction, temperature and emissivity, together with a systematic classification of candidate sources recovered from all-sky survey data. Combined with previous searches, including the warm circumstellar limits of \citet{suazo24}, the completed programme would establish constraints across nearly the full waste-heat temperature range, from several hundred kelvin to within a factor of a few of the CMB floor across the solar neighbourhood. Subject to the sensitivity limits of current facilities, this would constitute the first temperature-complete assessment of Dysonian technosignatures in the sixty-year history of the field. Future facilities can extend these limits to lower luminosities, greater distances and larger samples, but they would do so within a temperature domain that has finally become accessible across its full extent.

\subsection{Caveats} \label{sec:caveats} 

The migration argument is a suggestion rather than a theorem. Technological civilisations may not be computation-dominated, may prefer rapid operation at higher temperatures, may aestivate, or may be entirely absent. The search strategy developed here is therefore best viewed as a test of a particular thermodynamic hypothesis rather than a generic prediction of intelligent life favouring cold computing. The energy source is deliberately parametrised. The collector--radiator identity shows that local starlight is sufficient to power the structures considered here, but the search does not assume it. Systems with $\Ltech>L_{*}$ would instead point to alternative energy sources, including onboard generation, indirect energy extraction, or the import of energy from elsewhere via directed transmission (Section~\ref{sec:diagnostic}). Likewise, the mass budget assumes thin-film construction, but the dependence on material properties remains explicit in equation~(\ref{eq:mass}). 

The grey-body model is also an idealisation. Real structures may exhibit complex geometries, anisotropic emission, wavelength-dependent emissivities or non-thermal components. In particular, we assume similar absorption and emission efficiencies at the characteristic wavelengths of the stellar radiation field and the waste-heat spectrum. Departures from this assumption shift the equilibrium temperature and radius of a structure but do not alter the basic observables, which remain its luminosity and characteristic temperature. We also treat the cosmic microwave background as the ultimate external radiation bath. In practice, the interstellar radiation field raises the minimum equilibrium temperature to roughly 3.5--4~K within the Galactic disc, slightly narrowing the available low-temperature regime but not materially affecting the 5--60~K parameter space explored here. The discriminant based on spectral greyness assumes optically thick radiators. Artificial surfaces engineered to produce strongly wavelength-dependent emissivities could evade discriminant~(i), although they remain subject to the independent tests provided by discriminants~(v) and (vi). More generally, individual discriminants should not be regarded as decisive in isolation; the strength of the programme lies in their combined application. Observationally, the largest challenge is not sensitivity but interpretation. Candidate identification in the \textit{Planck} catalogues is complicated by the large beam size, which can make associations with individual stellar systems ambiguous and necessitates higher-resolution follow-up. Recovering smooth, arcminute-scale emission from the ground is also technically demanding. This difficulty is most acute for the largest halo candidates around nearby luminous stars and is much reduced for compact halo systems whose angular extent is comparable to a single beam or smaller. 

All these caveats caution against over-interpreting individual candidates, but they do not undermine the search logic itself, which rests on generic differences between engineered radiating surfaces and natural cold dust. Finally, any candidate surviving the discriminants will attract disproportionate attention; candidate handling should follow the recently updated SETI post-detection protocols  \citep{garrett25}.

\subsection{The local limit} 
\label{SScold}

The Slysh-halo argument applies equally to our own planetary system. If cold, dissipating artefacts exist in the outer Solar System, the same thermodynamics places them at hundreds of AU from the Sun. In this $d\rightarrow0$ limit, individual nodes that would be unresolved around other stars become detectable as discrete cold sources, while existing far-infrared and millimetre surveys already place constraints on their dissipated power. The Solar System therefore provides a useful local analogue of the circumstellar search problem considered here, extending the same physical framework from kiloparsec distances to our immediate cosmic neighbourhood.

\section{Conclusions}
\label{sec:conclusions}

The argument of this paper begins with an observation about the searches we
have already made and ends with a prescription for the ones we have not.
Sixty years of Dysonian SETI have been conducted almost entirely in the
mid-infrared and are therefore a search for technology that chooses to work
at 100--600~K. Nothing in physics privileges that choice; on the contrary, if
the long-term energy budget of a technological civilisation is dominated by
computation, then everything in physics argues against it. The Landauer cost
of an irreversible bit operation scales linearly with temperature,
refrigerating below ambient incurs a steep Carnot penalty, and every
planetary system supplies, for free, an
arbitrarily cold ambient environment at sufficiently large circumstellar
radius. A mature computational civilisation might therefore be highly motivated to migrate its facilities outward and down
the temperature ladder toward the CMB floor -- and the same physics that
makes this efficient makes it visible. For a given technological power budget, the required radiating area scales as $T^{-4}$, so cold computation is necessarily vast in extent and its waste heat emerges, undiminished, in the far-infrared and submillimetre.
Cold does not
mean faint; it means large. Radio silence is a choice; waste heat is not.

The observational expression of this argument is what we propose to call the \emph{Slysh halo}: a physically motivated realisation of a partial ($f\ll1$) Dyson swarm. It appears as extended, line-free thermal emission with a nearly grey spectrum ($\beta\approx0$), originating in the cold outer regions of planetary systems where equilibrium temperatures fall into the 5--30~K range. The structure remains associated with an otherwise normal stellar host and follows the equilibrium relation $T\propto r^{-1/2}$. Its appearance is reversed between wavebands: the halo is effectively invisible in the optical, while in the submillimetre the stellar photosphere contributes only a faint Rayleigh--Jeans tail beneath the halo emission. Nothing in the basic energetics or material requirements of such structures appears too extravagant on the scale of a very  advanced technological civilisation. Because a collector operating at the local equilibrium temperature requires collecting area comparable to its radiating area, ambient starlight is sufficient to power computation at virtually any circumstellar radius. Under the fiducial thin-film assumptions adopted here, a halo detectable in existing archival data at 10~pc requires only $\sim0.2$ Ceres masses of material, drawn from the substantial small-body reservoirs that planet formation naturally leaves in the outer reaches of planetary systems. In this framework, the classical enclosing Dyson shell survives only as the $f\rightarrow1$ limiting case. Yet such systems would themselves be conspicuous: a 10~K shell radiating $1\,\Lsun$ produces a flux density of $\sim335$~Jy at 545~GHz from 100~pc, making it detectable across a substantial fraction of the Galactic disc. Even around low-luminosity hosts, the reduced mass requirements partly offset the smaller luminosity scale, leaving complete cold shells well within the range of current and forthcoming submillimetre facilities.

Perhaps the most striking conclusion of this work is how much of the relevant observational material already exists. The DEBRIS, DUNES and SONS surveys, although designed as debris-disc programmes, contain the far-infrared and submillimetre measurements needed to constrain cold circumstellar waste heat around hundreds of nearby stars. Reinterpreted in this context, they can in principle be translated into constraints on compact Slysh-halo dissipation at levels of order $\sim10^{20}$~W ($f\sim10^{-7}$--$10^{-4}$), several orders of magnitude below published waste-heat limits in the mid-IR, while providing a direct route to mapping the combinations of halo luminosity, temperature and emissivity compatible with the data. These constraints are expected to be strongest for the nearby M dwarfs that dominate the local stellar population, whose compact halo zones are naturally matched to existing single-dish observations. Existing archives are therefore not merely a starting point for a future programme; they already probe an interesting fraction of the parameter space considered here. The same is true at larger angular scales and greater distances. Extended haloes have not been searched for systematically, yet they are potentially accessible through archival observations from \textit{Planck}, JCMT, ACT and SPT. At the opposite extreme, the compact angular scales of distant systems and the clumped structures expected in some technological architectures are naturally suited to interferometric observations, opening a complementary route through the reprocessing of archival ALMA and NOEMA data and future dedicated high-resolution follow-up. In all cases, the challenge is often not detection but interpretation: distinguishing candidate structures from cold dust, molecular clouds and other astrophysical contaminants. The discriminants assembled here, including spectral greyness, radial structure, environmental context, molecular-line follow-up and high-resolution imaging,  provide a practical framework for that task. A significant part of the search,  therefore consists not of acquiring new data, but of asking new questions of data that already exist.

What remains is to look where no one has looked before. Observations of the nearest stellar systems, especially the M dwarfs that dominate the local census, provide natural targets for dedicated submillimetre surveys capable of reaching fractional luminosities of order $f\sim10^{-6}$ with existing facilities. A complementary programme follows from the reprocessing of archival ALMA data and new interferometric observations, extending the search to compact, distant and potentially clumped structures that may escape detection in single-dish surveys. Together, these approaches transform the Slysh halo from a theoretical possibility into an observationally testable technosignature. Even the null result would be
of real value: combined with the mid-infrared limits of \citet{suazo24}, a
completed programme would close the waste-heat window from 600~K to within
a factor of a few of the CMB floor across the solar neighbourhood, the
first temperature-complete statement in the history of Dysonian SETI. The temperature of waste heat is not merely a spectral detail but a
physical statement about how technology operates. A temperature-complete
survey therefore probes not only the existence of technological activity
but also the strategies by which advanced civilisations manage energy,
computation, and entropy. A null result would eliminate one of the most physically
well-motivated pathways for mature technological activity, namely
large-scale cold computation powered by stellar energy. And
if the migration argument is right, the temperature axis is an age axis:
the warm searches sample technological adolescence, while the submillimetre
samples maturity, where long-lived populations accumulate. Our own civilisation, now sketching its first orbital, solar-powered data centres, has just set foot on the warm end of that axis. The coldest technosignature searches may offer our best shot at discovering the oldest and most advanced technological systems in our Galaxy. More remarkably still, the search can begin immediately, using observations that have already been made.

\section*{Acknowledgements}

We would like to thank the referee... 

[TBC.]

\section*{Data Availability}
All quantities derive from published catalogues and the analytic grey-body
model of Section~\ref{sec:model}; a short reproduction script (Python) that
generates the tables and figures is available on request.


\bsp
\label{lastpage}
\end{document}